\documentclass[twocolumn,showpacs,preprintnumbers,amsmath,amssymb, prl, reprint]{revtex4-1}
\usepackage{amsmath}
\usepackage{stmaryrd}
\usepackage{txfonts}
\usepackage{amssymb}
\usepackage{mathrsfs}
\usepackage{graphicx}
\usepackage{dcolumn}
\usepackage{bm}
\usepackage{epsfig}
\usepackage{color}
\usepackage[colorlinks=true, linkcolor=blue, citecolor=blue, urlcolor=blue]{hyperref} 
\begin{document}
\preprint{\href{http://dx.doi.org/10.1103/PhysRevB.93.060401}{S.-Z. Lin and A. Saxena , Phys. Rev. B {\bf 93}, 060401(R) (2016).}}

\title{Dynamics of Dirac strings and monopole-like excitations in chiral magnets under a current drive}
\author{Shi-Zeng Lin}
\email{szl@lanl.gov}
\affiliation{Theoretical Division, Los Alamos National Laboratory, Los Alamos, New Mexico 87545, USA}

\author{Avadh Saxena}
\affiliation{Theoretical Division, Los Alamos National Laboratory, Los Alamos, New Mexico
87545, USA}

\begin{abstract}
Skyrmion lines in metallic chiral magnets carry an emergent magnetic field experienced by the conduction electrons. The inflow and outflow of this field across a closed surface is not necessarily equal, thus it allows for the existence of emergent monopoles. One example is a segment of skyrmion line inside a crystal, where a monopole and antimonopole pair is connected by the emergent magnetic flux line. This is a realization of Dirac string-like excitations. Here we study the dynamics of monopoles in chiral magnets under an electric current. We show that in the process of creation of skyrmion lines, skyrmion line segments are first created via the proliferation of monopoles and antimonopoles. Then these line segments join and span the whole system through the annihilation of monopoles. The skyrmion lines are destroyed via the proliferation of monopoles and antimonopoles at high currents,  resulting in a chiral liquid phase.  We also propose to create the monopoles in a controlled way by applying an inhomogenous current to a crystal. Remarkably, an electric field component in the magnetic field direction proportional to the current squared in the low current region is induced by the motion of distorted skyrmion lines, in addition to the Hall and longitudinal voltage. The existence of monopoles can be inferred from transport or imaging measurements.
\end{abstract}
\date{\today}
\maketitle

Skyrmion in magnets is a stable mesoscopic topological excitation where spins wrap the surface of a sphere once.~\cite{Bogdanov89,Rosler2006} Skyrmions have been observed in magnetic materials, such as B20 chiral magnets without inversion symmetry ~\cite{Muhlbauer2009, Yu2010a,Yu2011}, multiferroics \cite{Seki2012,Adams2012} and  even centrosymmetric compounds \cite{Yu05062012,yu_biskyrmion_2014}. These findings suggest that skyrmions are ubiquitous in magnetic systems. The typical size of skyrmions is about 10 nm and skyrmions form a triangular lattice. Skyrmions can be driven by electric current ~\cite{Jonietz2010,Yu2012,Schulz2012}, temperature gradient ~\cite{Kong2013,Lin2014PRL,Mochizuki2014} and electric field gradient \cite{White2012,White2014}. Moreover the threshold current that makes skyrmions mobile is 5 to 6 orders of magnitude smaller than that for a magnetic domain wall. ~\cite{Jonietz2010,Yu2012,Schulz2012} Thus skyrmions are promising candidates for spintronics applications and recently they have attracted significant attention. ~\cite{Fert2013,nagaosa_topological_2013}

In metals, the spin of the conduction electron interacts with the localized magnetic moment associated with skyrmions through the Hund's coupling. In the strong coupling limit, the spin of a conduction electron changes adiabatically when it moves around a skyrmion and it gains a Berry phase which can be described by an effective magnetic field. The emergent magnetic field is related to the skyrmion topological charge, $\mathbf{B}^E=\hbar c\epsilon_{ijk}\mathbf{n}\cdot(\partial_j \mathbf{n}\times \partial_k \mathbf{n})/(2e)$, and is extremely strong due to the small size of the skyrmion. Here $\mathbf{n}$ is a unit vector representing the direction of localized spin and $\epsilon_{ijk}$ is the totally antisymmetric matrix with $i,\ j,\ k=x,\ y,\ z$. A skyrmion line in a 3D metal thus can be regarded as an emergent magnetic flux line with quantum flux $\Phi_0=hc/e$.  It was demonstrated experimentally that two skyrmion lines can merge and become a single line. This implies an emergent magnetic monopole with flux $\Phi_0$ at the merging point, which can be verified by integrating $\mathbf{B}^E$ over a closed surface around the merging point.~\cite{Milde2013} A monopole and antimonopole pair can also be created by thermal fluctuations. \cite{schutte_dynamics_2014} The emergent magnetic flux of the skyrmion and hence the monopole is an effective description for the conduction electrons \cite{Zang11}, thus these concepts cannot be applied to insulators. While magnetic monopoles as elementary particles have never been found experimentally, quasiparticles resembling monopoles have been identified in several condensed-matter systems, such as spin ice \cite{Castelnovo2008} and chiral magnets discussed here as well as Bose-Einstein condensates \cite{Ray01052015}. A unique feature of monopoles in skyrmion systems is that they can be driven by electric currents. The monopoles in chiral magnets have attracted considerable interest recently. \cite{Takashima2014,watanabe_electric_2014,Kawaguchi2015}

\begin{figure*}[t]
\psfig{figure=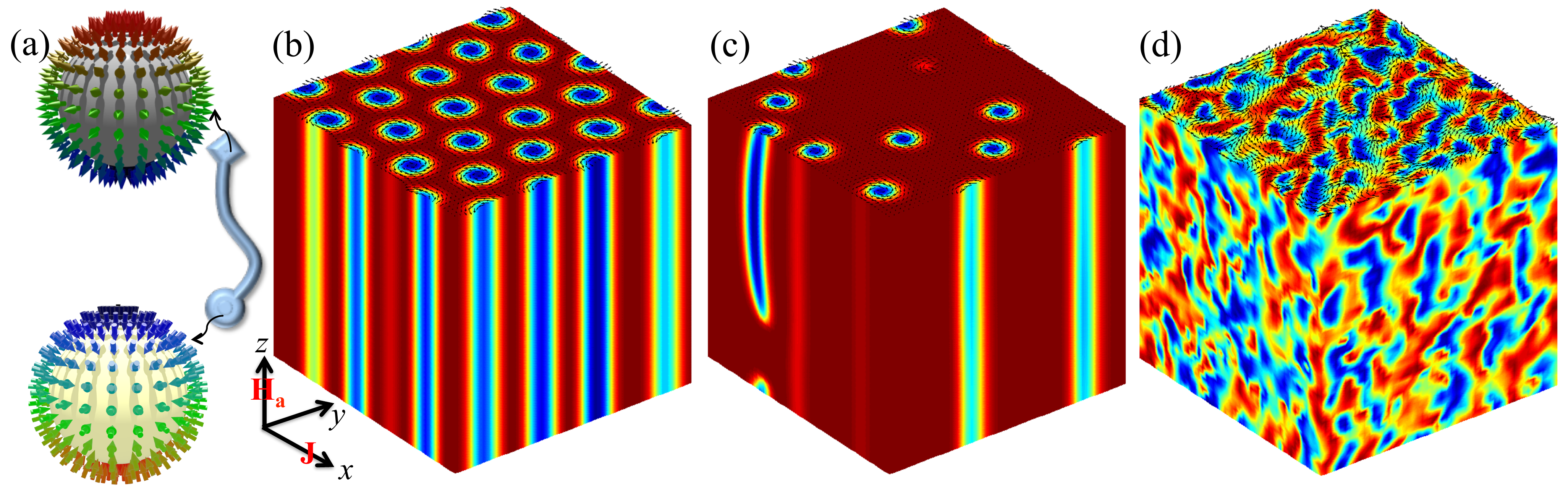,width=18cm}
\caption{(color online) (a) Schematic view of a skyrmion segment as a Dirac string-like excitation, where there is a monopole and antimonopole pair at the two ends. (b) Straight skyrmion line lattice. (c)  Snapshot of spin configuration after the magnonic instability of spin polarized state. Skyrmion segments are created dynamically associated with monopole and antimonopole excitations. The monopole and antimonopole annihilate or disappear at the surfaces, through the merging of the skyrmion segments. Finally the system reaches a state with straight skyrmion line lattice, and they move as a whole. (d) Chiral liquid at a high current when skyrmion lines are destroyed.  Color at three surfaces represents $n_z$ (red for $n_z=1$ and blue for $n_z=-1$) and the vector field at the top surface denotes the $n_x$ and $n_y$ components.
} \label{f1}
\end{figure*}

In a thin film, skyrmion becomes a pancake-like object and no monopole is allowed to appear because the system is uniform along the direction perpendicular to the film. It was demonstrated theoretically that skyrmions can be created dynamically by applying a strong current to a fully spin polarized state in chiral magnetic thin films. \cite{szlin13skyrmion1} This is because the current induces a Doppler shift in the magnon spectrum \cite{PhysRevB.69.174412,vlaminck_current-induced_2008} and renders the spin polarized state unstable when the spectrum becomes gapless, after which skyrmions are nucleated dynamically. When the current increases further, skyrmions are destroyed and the system evolves into a chiral liquid with strong spatial and temporal fluctuations in spin chirality.   

The creation and destruction of skyrmions by current can also occur in 3D, where the emergent monopoles play an important role. A skyrmion line percolating the whole system (connecting surfaces of the system) cannot be created instantly. Instead segments of skyrmion lines are created. At the ends of these lines, $\nabla\cdot \mathbf{B}^E=\pm \rho_m$, i.e. there is a monopole and antimonopole pair connected by a skyrmion line, as schematically shown in Fig. \ref{f1} (a). This is a realization of Dirac string in condensed matter systems. These segments then join together to form skyrmion lines spanning the whole system through the annihilation of monopoles and antimonopoles. In the course of skyrmion destruction, skyrmion lines break into segments and monopole and antimonopole pairs are created dynamically by current. These physical pictures are borne out by our numerical simulations below.

We consider a 3D chiral magnet described by the following Hamiltonian in the continuum limit because the skyrmion size is much bigger than the underlying lattice parameter \cite{Bogdanov89,Bogdanov94,Rosler2006,Han10,Rossler2011}
\begin{equation}\label{eq1}
\mathcal{H}=\int d\mathbf{r}^3 \left[\frac{J_{\rm{ex}}}{2}(\nabla \mathbf{n})^2-\mathbf{H}_a\cdot\mathbf{n} \right]+\mathcal{H}_{\mathrm{DM}},
\end{equation}
\begin{equation}\label{eq1a}
\mathcal{H}_{\mathrm{DM}}=D\int d\mathbf{r}^3 \left[n_x {\partial_y n_z}-n_y {\partial_x n_z}+n_z \left({\partial_x n_y}-{\partial_y n_x}\right)\right],
\end{equation}
where $J_{\rm{ex}}$ is the exchange coupling and $\mathcal{H}_{\mathrm{DM}}$ is the Dzyaloshinskii-Moriya (DM) interaction due to the absence of inversion symmetry. \cite{Dzyaloshinsky1958,Moriya60,Moriya60b}  Here we consider that the DM vector $\mathbf{D}$ is along the bond direction in the $x$-$y$ plane and the component along the $z$ direction is zero, $D_x=D_y=D$ and $D_z=0$. \footnote{ For $D_x=D_y=D_z=D$, the ordering vector of the helical phase is along the $[111]$ direction. Skyrmion lines are along the $[\bar{1}\bar{1}2]$ direction when the magnetic field is along the same direction. The magnitude and direction of $\mathbf{D}$ do not change the physics discussed here qualitatively.} The field is along the $z$ direction, $\mathbf{H}_a=H_a\hat{z}$ with a unit vector $\hat{z}$,  and the skyrmion lines are aligned in the same direction. In 3D systems, skyrmions are stable only in a small phase region in magnetic field-temperature parameter space \cite{Muhlbauer2009,Buhrandt2013}. Here we consider a metastable skyrmion line lattice at zero temperature which can be obtained through field cooling \cite{Milde2013}. The dynamics of spins follows the Landau-Lifshitz-Gilbert equation
\begin{equation}\label{eq2}
{\partial _t}{\bf{n}} = \frac{\hbar\gamma}{2e}({{\bf{J}} }\cdot\nabla) {\bf{n}} - \gamma {\bf{n}} \times {{\bf{H}}_{\rm{eff}}} + \alpha {\bf{n}}\times  {\partial _t}{\bf{n}} ,
\end{equation}
where $\gamma$ is the gyromagnetic ratio, $\alpha$ is the Gilbert damping constant, $\mathbf{H}_{\rm{eff}}\equiv-\delta \mathcal{H}/\delta{\mathbf{n}}$ is the effective magnetic field and $\mathbf{J}$ is the spin polarized current. The current is in the $x$ direction in our simulations.

 \begin{figure*}[t]
\psfig{figure=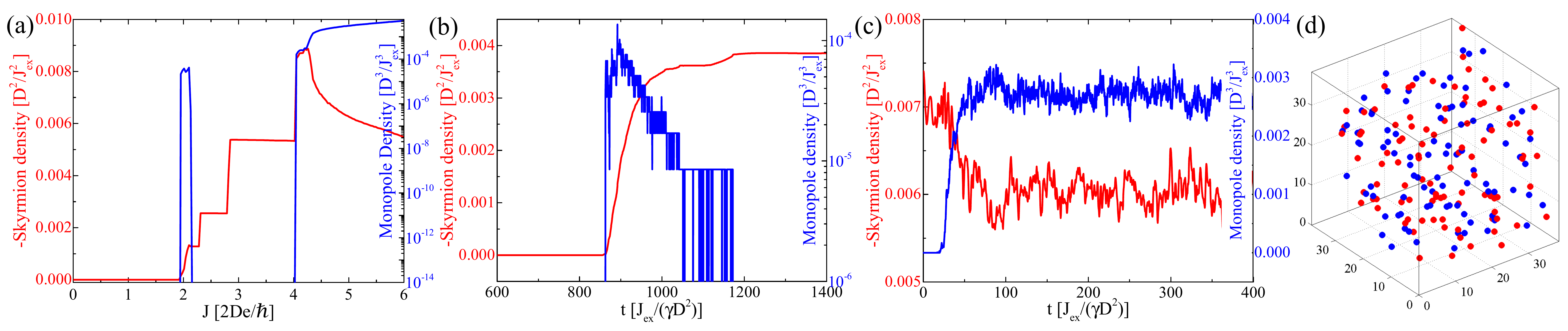,width=18cm}
\caption{(color online) (a) Skyrmion and monopole density versus current in the stationary state for $H_a=0.8\ D^2/J_{\mathrm{ex}}$. At $J=3.84\ De/\hbar$, skyrmion segments are created after the instability of the ferromagnetic state. (b) Time dependence of skyrmion and monopole density when the instability is triggered at $J=3.7\ De/\hbar$ at $t=0$ for $H_a=0.6\ D^2/J_{\mathrm{ex}}$. Monopoles and antimonopoles are created after the instability but they finally disappear in the stationary state, resulting in a straight skyrmion line lattice. (c) Dynamics of skyrmion destruction at a high current $J=10.0\ De/\hbar$ and $H_a=0.5\ D^2/J_{\mathrm{ex}}$, where monopoles and antimonopoles are nucleated. (d) Distribution of monopoles and antimonopoles in the chiral liquid phase at $J=8.64\ De/\hbar$ and $H_a=0.8\ D^2/J_{\mathrm{ex}}$.} \label{f2}
\end{figure*}

In simulations, we calculate the skyrmion density in the $x$-$y$ plane $\rho(\mathbf{r})=\mathbf{n}\cdot(\partial_x\mathbf{n}\times\partial_y\mathbf{n})/(4\pi)$, or the $B_z^E$ component. The emergent electric field induced by skyrmion motion is $\mathbf{E}^E=\hbar\mathbf{n}\cdot(\nabla\mathbf{n}\times\partial_t\mathbf{n})/(2e)$. The system size is $40\times40\times40\ (J_{\mathrm{ex}}/D)^3$. The system is discretized into a cubic mesh with periodic boundary conditions in most cases. To count the monopole/antimonopole number, we split each surface of a unit cubic cell into two triangles, and calculate the solid angle $\Omega_i$ subtended by the three spins $\mathbf{n}_i$ in a triangle, $\tan(\Omega_i/2)=\mathbf{n}_1\cdot(\mathbf{n}_2\times\mathbf{n}_3)[1+\mathbf{n}_1\cdot\mathbf{n}_2+\mathbf{n}_2\cdot\mathbf{n}_3+\mathbf{n}_3\cdot\mathbf{n}_1]^{-1}$. \cite{Milde2013} The monopole number of this unit cell is obtained by summing the solid angle for these 12 triangles on the surface of the unit cubic cell, i.e. $n_M=(4\pi)^{-1}\sum_{i=1}^{12}\Omega_i$, which is quantized, $n_M=\pm 1,\ 0$.

We first study the nucleation of skyrmions in the ferromagnetic state due to the magnon instability induced by current. The magnon dispersion in the fully spin polarized state is given by \cite{szlin13skyrmion1}
\begin{equation}\label{eq3}
\Omega(\mathbf{k})  =\frac{\hbar\gamma}{2e} \mathbf{J}\cdot \mathbf{k}+ \frac{\gamma (1+i\alpha) }{\alpha ^2+1}\left(H_a+ J_{\rm{ex}} \mathbf{k}^2\right).
\end{equation}
The term $\mathbf{J}\cdot \mathbf{k}$ accounts for the magnonic Doppler shift. The magnon gap for $J=0$ is due to the external field and it becomes gapless at a threshold current $J_m={4e \sqrt{H_a J_{\text{ex}}}}/[\hbar({\alpha ^2+1})]$, signaling an instability. The averaged skyrmion and monopole density in the stationary state as a function of current is displayed in Fig. \ref{f2} (a). Right after the instability, skyrmion segments are nucleated accompanying the creation of monopoles and antimonopoles, see Fig. \ref{f1} (c). In Fig. \ref{f2} (b), the skyrmion and monopole densities as a function of time are shown when the instability is triggered. Short skyrmion segments appear with a high monopole density. During the relaxation, the length of skyrmion segment increases accompanying an increase in skyrmion density. Once these segments meet, monopole and antimonopole annihilate resulting in longer skyrmion segments. This process repeats until the skyrmion lines span the whole system when no monopole and antimonopole is left. Meanwhile the skyrmion lines become straight to minimize the line tension. The monopole density decreases exponentially and the typical time scale for this process is about $300\ J_{\mathrm{ex}}/(\gamma D^2)$. In the end, straight skyrmion line lattice driven by current moves as a whole, see Fig. \ref{f1} (b). 

The skyrmion density increases stepwise with current. For a low skyrmion density, the system is close to a ferromagnetic state and one would expect a similar magnonic instability as that in Eq. \eqref{eq3}. For $J$ close to $J_m$, the skyrmion density is low and the magnetic instability persists at the same current. In this case skyrmion segments do not relax into straight lines. When the skyrmion density becomes high, the instability current increases. The magnonic instability disappears when the skyrmion lattice is reached at a high skyrmion density for an intermediate $J$. This accounts for the stepwise increase of skyrmion density. In these steps, no monopole exists because the skyrmion line lattice spans the whole system.

At a high current, the skyrmion density decreases sharply indicating destruction of skyrmions [see Fig. \ref{f2} (a)]. The system evolves into a chiral liquid displayed in Fig. \ref{f1} (d), where spin chirality fluctuates strongly in space and time. In the course of skyrmion destruction, skyrmion lines break into segments with a monopole and an antimonopole at ends (similar to the Dirac strings). In other words, skyrmion lines are destroyed due to the proliferation of monopole-antimonopole pairs in this dynamic phase transition. The time evolution of monopole and skyrmion density is presented in Fig. \ref{f2} (c). Skyrmion density decreases while the monopole density increases. In the chiral liquid, both densities fluctuate justifying the liquid nature of the final state. These fluctuations are induced dynamically by current. As shown in Fig. \ref{f2} (d),  the monopoles  and antimonopoles are distributed randomly in space, and can be regarded as a monopole gas. The dynamical phase transitions associated with the skyrmion creation and destruction are of the first-order because there is a strong hysteresis. \cite{szlin13skyrmion1} Both transitions can be detected from current-voltage characteristics because the emergent electric field depends on the skyrmion density.

\begin{figure*}[t]
\psfig{figure=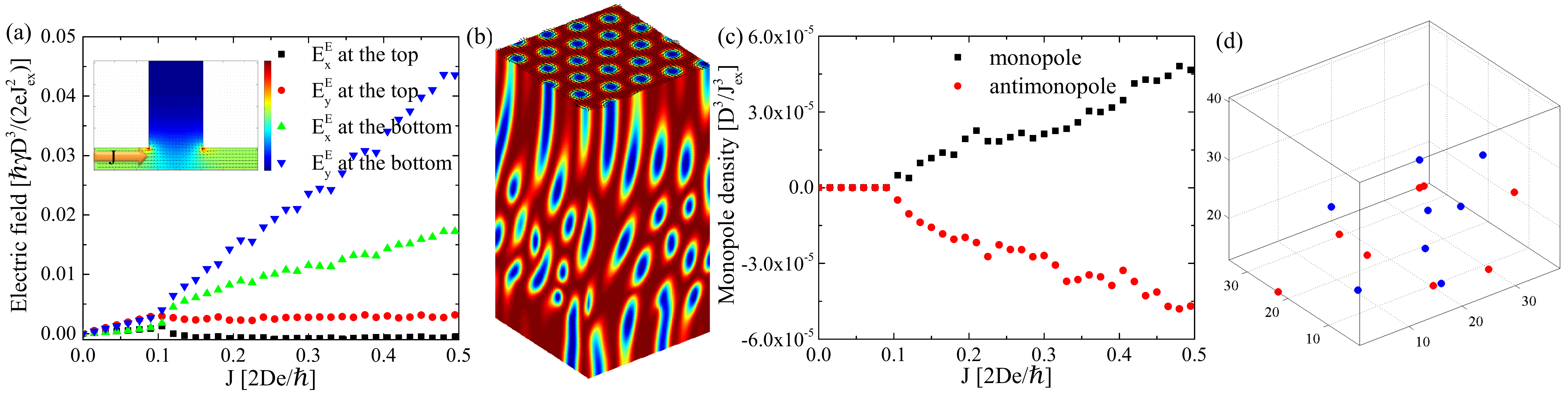,width=18cm}
\caption{(color online) (a) Emergent electric fields at the top and bottom surfaces of the crystal. At a small current, $\mathbf{E}^E$ at the opposite surfaces is the same indicating an elastic motion of skyrmion lines. Beyond a threshold current, $\mathbf{E}^E$ at the bottom surface under a current drive is much bigger than the top one, implying the breaking of skyrmion lines. Inset is a sketch showing how to inject current at the bottom surface and the calculated current distribution \cite{supplement}. (b) Snapshot of spin configuration after the breaking of skyrmion lines.  (c) Monopole and antimonopole density versus current. (d) Distribution of monopoles and antimonopoles after the segmentation of skyrmion lines. In this calculation, we have used open boundary conditions in the $z$ direction. Here $H_a=0.5\ D^2/J_{\mathrm{ex}}$ and $J=0.26\ De/\hbar$ for (b) and (d). The system size is $L_z=2L_x=2L_y=100\ {J_{\mathrm{ex}}/D}$.
} \label{f3}
\end{figure*}

We next discuss the controlled creation of monopoles and antimonopoles by driving part of the skyrmion lines. For ease of discussion, let us consider the case that current only flows at the bottom surface of the crystal. The skyrmion lines near the surface are driven by the Lorentz force due to the current, $\mathbf{J}\delta(z)$, and start to move. This causes distortion of the lines and drags the skyrmion lines inside the crystal to move due to the elastic energy of the skyrmion lines [see Fig. \ref{f3} (b)]. For a small current, the elastic force can balance the different Lorentz forces acting on the surface and middle of the skyrmion line, and the skyrmion lines move as a whole. For an elastic skyrmion line lattice, the equation of motion for the lattice can be described by a displacement vector $\mathbf{u}(z)=[{u}_x(z),\ u_y(z),\ 0]$ \cite{szlin13skyrmion2,Iwasaki2013}
\begin{equation}\label{eq4}
\frac{4\pi}{\gamma}\left[\eta {\mathbf{v}}(z)+\mathbf{v}(z)\times\hat z\right]=\frac{\Phi_0}{c}\hat{z}\times \mathbf{J}\delta(z)+\frac{\ell}{2}\partial_z^2 \mathbf{u}(z),
\end{equation}
where $\mathbf{v}=[\partial_t u_x,\ \partial_t u_y,\ 0]$ is the skyrmion velocity, and $\eta\ll 1$ is the dimensionless viscosity due to the Gilbert damping for localized spins and the Ohmic dissipation by conduction electrons. Here $\ell$ is the tilt modulus of the skyrmion lattice and it becomes skyrmion self-energy per unit length for a single line. From Eq. \eqref{eq4}, the velocity component parallel to the current is $v_{\parallel}=-\gamma \hbar J/[2 e(1+\eta^2)L_z]$ and that perpendicular to the current is $v_{\perp}=\gamma \hbar \eta J/[2 e(1+\eta^2)L_z]$ with $L_z$ the system size along the $z$ direction. For a small current, the equation of motion is the  same as that for a single skyrmion in thin films. The displacement field is $u\propto J z^2$.

 \begin{figure}[b]
\psfig{figure=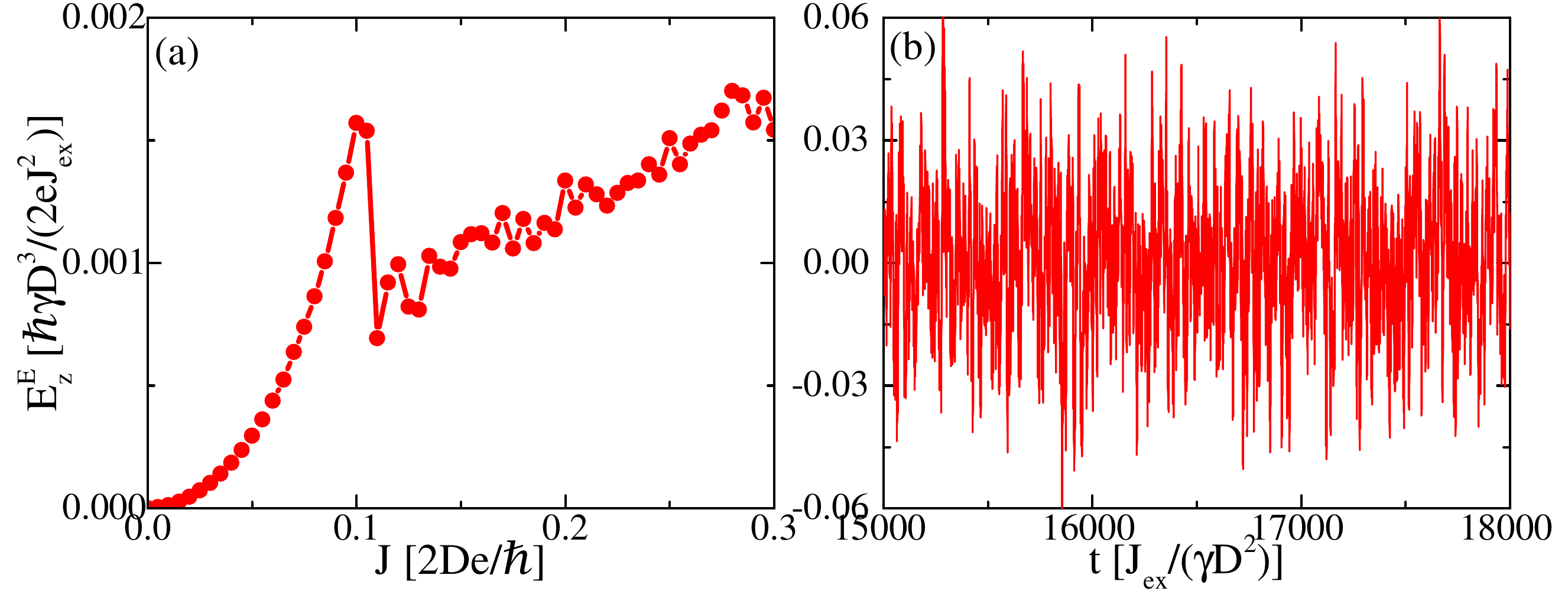,width=\columnwidth}
\caption{(color online) (a) Emergent electric field along the magnetic field direction $E_z^E$ when skyrmion lines are stretched by a current at the bottom surface in a T-shaped crystal in Fig. \ref{f3} (a). (b) Fluctuations of $E_z^E$  as a function of time in the chiral liquid phase at $J=10.0\ De/\hbar$. Here $H_a=0.5 \ D^2/J_{\mathrm{ex}}$. } \label{f4}
\end{figure}

However for a strong current, the skyrmion line breaks. The skyrmion segments at the surface and inside the crystal travel with different velocities. This decoupled motion of the skyrmion segments is possible because the monopole and antimonopole pairs are allowed for the emergent magnetic flux line associated with the skyrmion. 

The above picture is confirmed by our simulations. We consider a T-shaped crystal shown in the inset of Fig. \ref{f3} (a). The current is injected through the bottom of the crystal. The skyrmion lines produce an emergent electromagnetic field which affects the motion of conduction electrons. However, as shown in the Supplement Materials \cite{supplement}, the emergent electric field is much smaller than the real electric field induced by the injected  current, and we neglect the emergent electric field in the calculation of current distribution. The current distribution is obtained by solving $\nabla^2 V=0$, subject to the proper boundary conditions, with $V$ being the voltage \cite{supplement}. As displayed in Fig. \ref{f3} (a), below a threshold current, the electric field at the top and bottom are the same, indicating motion of skyrmion lines as a whole. Beyond the threshold current, the skyrmion segments at the bottom surface move at a higher velocity. At the same time, monopoles and antimonopoles are created which are confined at the bottom surface [see Figs. \ref{f3} (c) and (d)]. At a large current, the skyrmion density near the bottom surface decreases due to the dynamical destruction of skyrmions. This causes a rapid increase of antimonopole density. Meanwhile the skyrmion density in the current-free region does not change, which results in a higher density of  antimonopoles than that of monopoles, see Fig. \ref{f3} (c).

The electric field along the magnetic field direction $E_z^E$ is particularly interesting. Let us first consider $E_z^E$ due to the stretching of skyrmion lines by a surface current. The electric field is given by $\mathbf{E}^E={\mathbf{v}}\times\mathbf{B}^E/c$. For a straight skyrmion line, $E_z^E=0$ because $\mathbf{B}^E=B^E\hat{z}$ only has the $z$ component. When the skyrmion lines are distorted according to Eq. \eqref{eq4}, $\mathbf{B}^E$ acquires an in-plane component, $\mathbf{B}^E=B^E[\partial_z u_x,\ \partial_z u_y,\ 1]/\sqrt{(\partial_z u_x)^2+(\partial_z u_y)^2+1}$   and $E_z^E$ is nonzero, as shown in Fig.~\ref{f4} (a). The induced electric field $E_z^E\propto J^2$ increases \emph{nonlinearly} with current in the elastic region and it suddenly decreases when the skyrmion line breaks, after which skyrmion segments become mostly aligned with the magnetic field.  In the chiral liquid phase, $E_z^E$ fluctuates around zero due to the fluctuations of the density of monopoles and antimonopoles; see Fig. \ref{f4} (b). Therefore $E_z^E$ provides a clear transport signature of the existence of monopoles. We remark that the nonuniform current distribution can also produce a local real electric field $E_z$ along the magnetic field direction. However $E_z$ averaged over the crystal vanishes because there is no net current in the magnetic field direction. 

The existence of monopoles can also be inferred from imaging measurements, such as magnetic force microscope \cite{Milde2013}. For instance, one may first apply a strong current to drive the system into the chiral liquid phase. Then the current is removed and the system can be trapped in a metastable state with monopoles. By visualizing the skyrmion density at the opposite surface, one can extract the information about the monopoles inside the crystal.

If we start with a skyrmion line lattice in the ground state and drive it, the magnon instability due to the Doppler shift is absent. At a high current, we still have the destruction of skyrmion lines in nonequilibrium similar to those in Fig. \ref{f2} (a) via the proliferation of monopoles and antimonopoles.

To summarize, we show that the emergent monopoles play an important role in the skyrmion line creation and destruction process in the presence of an electric current. The current density required for the monopole proliferation is of the order of $10^{12}\ \mathrm{A/m^2}$ for typical material parameters. The monopoles can also be created in a controlled way by applying an inhomogeneous current to the crystal. Especially we predict an induced electric field component along the magnetic field direction, in addition to the Hall and longitudinal electric fields.

\noindent \textit{Acknowledgments --}
The authors are indebted to Achim Rosch and Jiadong Zang for helpful discussions. The authors also thank Cristian D. Batista for helpful suggestions and a critical reading of the manuscript. Computer resources for numerical calculations were supported by the Institutional Computing Program at LANL. This work was carried out under the auspices of the National Nuclear Security Administration of the US DOE at LANL under Contract No. DE-AC52-06NA25396 and was supported by the LANL LDRD-DR Program.

\bibliography{reference}

\newpage
\appendix

\section{Calculation of the current distribution in a T-shaped crystal}
 \begin{figure}[t]
\psfig{figure=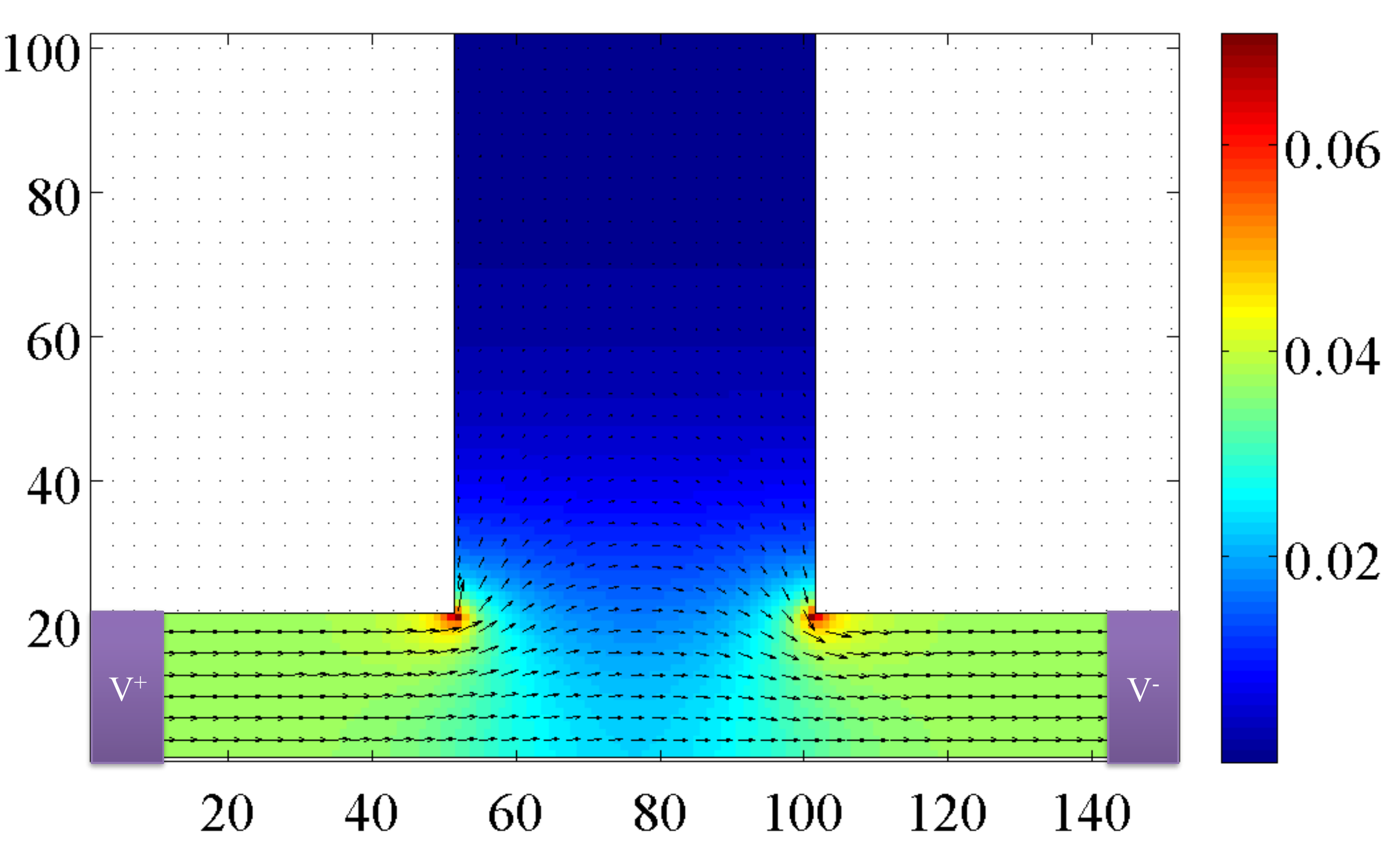,width=\columnwidth}
\caption{(color online) T-shaped crystal and the resulting current distribution. The vector field denotes the $x$ and $y$ components of $\mathbf{J}$ and color represents the intensity.} \label{fs1}
\end{figure}
We consider a T-shaped crystal sketched in Fig. \ref{fs1}. A current is applied into the crystal through two electrodes at the bottom, $V^+$ and $V^-$. The presence of skyrmion affects the motion of conduction electrons and changes the conductivity of the crystal. The skyrmion modifies the conductivity in two different ways. The skyrmion produces an effective magnetic field acting on the conduction electrons and changes the conductivity. Meanwhile when the skyrmions start to move, the motion of skyrmions produces an emergent electric field which also modifies the conductivity. For the transverse conductivity, the former contribution is known as the topological Hall conductivity and the latter as the skyrmion flow conductivity.  For MnSi, both contributions due to the skyrmions are found to be small compared to the background conductivity from the conduction electrons in experiments. \cite{Neubauer2009,Schulz2012} Below we provide an explicit estimate of the emergent electric field induced by skyrmion motion and compare it to the real electric field. The emergent electric field is \cite{Zang11}
\begin{equation}
\mathbf{E}^E\approx \frac{h \mathbf{v}}{e a_s^2},
\end{equation}
where $a_s$ is the linear size of skyrmion and $\mathbf{v}$ is the skyrmion velocity, which can be estimated using the Thiele's equation of motion for skyrmions \cite{szlin13skyrmion2}
\begin{equation}
\mathbf{v}\approx \frac{a^3\mathbf{J}}{2e},
\end{equation}
with $a$ the lattice constant for the spin system. The ratio of $\mathbf{E}^E$ to the real electric field $\mathbf{E}=\rho\mathbf{J}$ with $\rho$ being the resistivity is
\begin{equation}
\frac{\mathbf{E}^E}{\mathbf{E}}\approx\frac{h a^3}{2e^2 a_s^2\rho}.
\end{equation}
For MnSi, we have $a_s\approx 10$ nm, $a\approx 0.45$ nm and $\rho\approx 3\times 10^{-7}\ \mathrm{\Omega\cdot m}$ \cite{PhysRevB.82.155124}, thus we estimate ${\mathbf{E}^E}/{\mathbf{E}}\approx 0.03$. Therefore, we can neglect the skyrmion contribution in the calculations of the electric current distribution in the crystal below.

The current distribution is governed by the equation for voltage $V$
\begin{equation}\label{eqss1}
\nabla^2 V=0.
\end{equation}
The boundary condition in the region without electrodes is that the spatial derivative of the voltage normal to the boundary vanishes, $\partial_\mathbf{l} V=0$. For the boundary in contact with electrodes, the boundary condition is $V=V^+$ and $V=V^-=0$. The resulting current distribution is $\mathbf{J}=-\sigma \nabla V$, with $\sigma$ being the conductivity. We solve Eq. \eqref{eqss1} numerically for the geometry shown in Fig. \ref{fs1}.

The current mainly flows in the bottom region of the crystal. The current distribution does not depend on the intensity of the current because we can obtain the desired current intensity simply by rescaling $V^+$. The region where current extends in the vertical direction is of the same order as the lateral dimension of the crystal at the top. The nonuniform electric current distribution yields a local electric field in the vertical direction $E_z$. However, the averaged $E_z$ over the crystal vanishes, $\int dr^3 E_z=\int dr^3 J_z/\sigma=0$, because there is no net electric current in the vertical direction.

Knowing the current distribution in the crystal, we then solve the Landau-Lifshitz-Gilbert equation numerically.

\end{document}